\documentstyle[multicol,aps,prb]{revtex}

\begin{document}

\draft

\title{
Ferromagnetism in a Hubbard model for an atomic quantum wire:\\
a realization of flat-band magnetism from even-membered rings
}

\author{
Ryotaro Arita, Kazuhiko Kuroki, Hideo Aoki, Akio Yajima and Masaru Tsukada
}
\address{Department of Physics, University of Tokyo, Hongo,
Tokyo 113, Japan}

\author{
Satoshi Watanabe
}
\address{Department of Materials Science, University of Tokyo, Hongo,
Tokyo 113, Japan}

\author{
Masahiko Ichimura, Toshiyuki Onogi and Tomihiro Hashizume
}
\address{Advanced Research Laboratory, Hitachi, Ltd., Hatoyama, Saitama
350-03, Japan}

\date{\today}

\maketitle

\begin{abstract}
We have examined a Hubbard model on a chain of squares,
which was proposed by Yajima {\it et al} as a model of an atomic
quantum wire As/Si(100),
to show that the flat-band ferromagnetism 
according to a kind of  Mielke-Tasaki mechanism should be realized 
for an appropriate band filling
in such a non-frustrated lattice.  
Reflecting the fact that the flat band is not a bottom one, 
the ferromagnetism vanishes, rather than intensified, 
as the Hubbard $U$ is increased.  
The exact diagonalization method is used to 
show that the critical value of $U$ 
is in a realistic range.
We also discussed the robustness of
the magnetism against the degradation 
of the flatness of the band.  

\end{abstract}

%\medskip


\begin{multicols}{2}
\narrowtext

Recent progress in atom manipulation 
on solid surfaces using scanning tunneling microscope
has opened up an unprecedented possibilities for 
designed structures on atomic scales. 
Specifically, Hashizume {\it et al}\cite{Hashizume} has 
fabricated
atomic wires from Ga adatoms bound to a row of 
dangling bonds on a hydrogen-terminated Si(100)2$\times$1 surface. 
Watanabe {\it et al} \cite{Watanabe} have 
obtained its band structure with the first-principles 
calculation.  Interestingly, they have found 
that a partially flat band appears 
in the dispersion and suggested a possibility of the ferromagnetism 
arising from the flat band.  
Some of the present authors \cite{Yajima} have then 
extended the electronic-structure calculation by going from 
a group III adatom (Ga) to a group V adatom 
(As).  
Unexpectedly, a flat (dispersionless) band appears 
for some atomic configurations for the As wire as well, which again 
suggests a possibility of ferromagnetism.  
As a possible origin of the flat band, Yajima {\it et al} \cite{Yajima}
have proposed a 
tight-binding model, which is thought to capture 
the essential feature of the 
arrangement of the arsenic atoms on the 
wire as a chain of connected squares.(fig.\ref{yajimodel})

Possibility of ferromagnetism due to the electron-electron 
repulsion on flat one-electron bands, or 
a macroscopic degeneracy in the one-electron states, 
was first shown rigorously for the Hubbard model by 
Lieb\cite{lieb}.  
While Lieb considered the models on bipartite lattices that have 
different
numbers of sublattice sites ($N_A - N_B \propto$ the magnetization), 
Mielke\cite{Mielkem1,Mielkem2} 
and Tasaki\cite{Tasakim1} have later considered 
line-graph theoretical (non-bipartite) models in which flat bands 
appear from interferences between
the nearest-neighbor transfer $t$ and more distant transfer $t'$.
The interference in each plaquette (which is triangular) 
is in fact a key to their 
rigorous reasoning why the flat band, when half-filled, 
leads to the ferromagnetism 
in the presence of the Hubbard repulsion $U$.
One remarkable feature of the Mielke-Tasaki theorem is that 
an infinitesimal $U$ is enough to make the ground state 
fully spin-polarized despite the fact that the magnetism 
appear from the electron correlation.

Now, Yajima's model for the atomic wire is bipartite, 
so that Lieb's theorem is applicable if the site energies 
are all identical.  Since the model is a chain 
of squares with $N_A = N_B$, 
the ground state is spin-singlet at half-filling (number of electrons 
= number of sites).  
Hence, 
a kind of Mielke-Tasaki mechanism has to 
somehow work to realize ferromagnetism in this model.

Usually Mielke's models (as exemplified by Kagom\'{e} lattice) 
and Tasaki's models are conceived as the graphs comprising triangles.
Penc {\it et al}\cite{Penc} have actually shown that 
a chain of connected triangles is favorable for ferromagnetism.
One is then tempted to interpret the interference 
mentioned above as an outcome of a frustration in 
odd-membered rings.  
So the challenge here is that: 
is it possible to have ferromagnetism \`{a} la Mielke-Tasaki 
for non-frustrated (e.g., a chain of even-membered rings as 
in the model proposed by Yajima {\it et al}).  
This is a doubly nontrivial question, 
since, even when one has a flat band from even-membered rings, 
it is by no means a sufficient condition for ferromagnetism and 
the applicability of the Mielke-Tasaki mechanism has to be examined.  
In this paper, we address to this question 
to conclude that, unexpectedly, 
a ferromagnetism \`{a} la Mielke-Tasaki does indeed exist 
for the non-frustrated model for the 
atomic-scale wire structure.

We consider the Hamiltonian, which consists of the one-electron 
tight-binding part, ${\cal H}_0$, and the Hubbard repulsion, 
${\cal H}_U$, where
\begin{eqnarray*}
{\cal H}_0&=&\sum_{i=1}^L \sum_{\sigma} \left[ t_1(
d^{\dagger}_{i+1,\sigma}a_{i \sigma} 
+a_{i \sigma}^{\dagger}d_{i \sigma})
+t_2 b_{i \sigma}^{\dagger}c_{i \sigma}+
\right. \\
&&\left. 
+t_3(
a^{\dagger}_{i \sigma}b_{i \sigma}
+c_{i \sigma}^{\dagger}d_{i \sigma} 
)+{\rm h.c.} \right]\\
&&+\sum_{i=1}^{L} \sum_{\sigma}\Delta \varepsilon(
n^{b}_{i\sigma}+n^{c}_{i\sigma}) , \\
{\cal H}_U&=&U\sum_{i=1}^{L} (n^{a}_{i\uparrow}n^a_{i\downarrow}+
n^{b}_{i\uparrow}n^{b}_{i\downarrow}+n^{c}_{i\uparrow}
n^{c}_{i\downarrow}
+n^{d}_{i\uparrow}n^{d}_{i\downarrow}).
\end{eqnarray*}
Here $a^{\dagger}_{i\sigma}\sim d^{\dagger}_{i\sigma}$ create, 
respectively, electrons with spin $\sigma$ at the four inequivalent 
positions a $\sim$ d on the square (fig.\ref{yajimodel}) 
in the $i$th unit cell, 
while 
$t_1$ is transfer between Si, $t_2$ between As,
and $t_3$ between Si and As.  
We have assumed a level offset, $\Delta \varepsilon$ 
between the As level and the Si level with 
$n^{b}_{i\sigma} \equiv b^{\dagger}_{i\sigma}b_{i\sigma}$ etc 
being number operators.  
Here we adopt periodic boundary conditions for the system 
having $L$ unit cells, and we take $t_1=1$.

This model has a special feature that, 
although the lattice contains only even-membered rings,
we do have flat band(s) if the parametric relation, 
\begin{eqnarray*}
t_1&=&t_2\\
t_3&=&t_1 \cos\theta\\
\Delta \varepsilon&=&t_1 (-1-2\sin\theta)
\end{eqnarray*}
is fulfilled, where $\theta$ is arbitrary 
(except for the condition 
$\theta\neq -\pi/2,0$ as discussed later).  
It is in fact straightforward to show that 
the single-electron Schr\"{o}dinger
equation has $L$-fold degenerate solutions 
with the eigenenergies
\begin{equation}
\varepsilon_{0} = -1-\sin\theta .
\label{eigenvalue}
\end{equation}

Namely, this model has in general one flat band 
(the second from the bottom) for a given value of $\theta$.  
We can alternatively place the flat band the third from the bottom 
($\varepsilon_{0} = 1-\sin\theta$) with 
$\Delta \varepsilon=t_1 (1-2\sin\theta)$, which is 
equivalent to the former through gauge and electron-hole transformations. 
If one wishes to have $\Delta \varepsilon =0$, 
then one can set $\theta=- \pi /6$, which is a special case of 
two flat bands (at $\varepsilon=\pm 1/2$) with an electron-hole symmetry.  
In fig.\ref{band} we show typical band structures.  
The fact that the structure is constructed from 
even-membered rings has a reflection that, 
unlike Mielke's model or Tasaki's model, 
the flat band is not the bottom band, but a dispersive 
band lies below (unless one changes the sign of 
one of the transfers in the ring).  

Our first key finding is that 
the reasoning employed by Mielke and Tasaki in ref[9]
can be applied to the 
present flat band.  
We can anticipate this, since the eigenfunctions on the flat band,
\begin{eqnarray*}
\phi_{i \sigma}&\equiv&\tilde{\phi}_{i \sigma}^{\dagger}|0 \rangle\\
&=&[b_{i\sigma}^{\dagger}+\sin\theta \, c_{i\sigma}^{\dagger}
-\cos\theta \, d_{i\sigma}^{\dagger} \\
&&+\cos\theta \, a^{\dagger}_{i+1\sigma}
- \sin\theta \, b^{\dagger}_{i+1\sigma}
- c^{\dagger}_{i+1\sigma}]|0 \rangle ,
\end{eqnarray*}
have a 
notable property that each eigenfunction cannot be accommodated 
within a unit cell of the structure (a shaded area in fig.1).  
This feature is favorable for a realization of the 
connectivity condition conceived by Mielke \cite{Mielkeg1}.  

We first apply the reasoning employed by
Mielke and Tasaki\cite{MieTasa} to the flat band here, 
while the effect of other bands are examined later.  
For this we can use the subspace spanned by
$\{\phi_{i,\sigma}\}$ for the number of 
electrons fixed to $L$ (i.e., the half-filled flat band).  
Any states in this subspace are represented as
\[
\Phi=\sum_{|L_{\uparrow}|+|L_{\downarrow}|=L}
f(L_{\uparrow},L_{\downarrow}) \prod_{i \in L{\uparrow}}
\tilde{\phi}_{i \uparrow}^{\dagger}
\prod_{j \in L{\downarrow}}
\tilde{\phi}_{j \downarrow}^{\dagger} |0\rangle
\]
where $L_{\uparrow}, L_{\downarrow}$ are arbitrary subsets of
$\{1\cdots L\}$ with 
$|L_{\uparrow}|$ and $|L_{\downarrow}|$ being the number of 
elements, respectively.

If there is a state that satisfies ${\cal H}_U \Phi_{GS}=0$, then
$\Phi_{GS}$ is the ground state in this subspace and, 
following the argument by Mielke and Tasaki\cite{MieTasa},
we can show for any $U>0$ that such states 
exist, non-degenerate apart from the trivial spin degeneracy
as follows.  
First we require that $n^{a}_{i \uparrow}n^{a}_{i \downarrow} \Phi_{GS}=0$
or $n^{d}_{i \uparrow} n^{d}_{i \downarrow} \Phi_{GS}=0$
for any $i$. 
To satisfy this condition, we must choose $L_{\uparrow}$ and
$L_{\downarrow}$ so that $L_{\uparrow}\cap L_{\downarrow}= \emptyset$.
Hence, we can represent any ground state as
\[
\Phi_{GS}=\sum_{\tilde{\sigma}} g(\tilde{\sigma})\prod_{i}
\tilde{\phi}^{\dagger}_{i \tilde{\sigma}(i)} |0\rangle
\]
where $\tilde{\sigma}$ represents the spin configuration.

For ${\cal H}_U \Phi_{GS}=0$ to be fulfilled 
we require $n^{b}_{i \uparrow}n^{b}_{i \downarrow} \Phi_{GS}=0$
or $n^{c}_{i \uparrow}n^{c}_{i \downarrow} \Phi_{GS}=0$ 
for any $i$.  This requirement amounts to 
\begin{eqnarray*}
0 = b_{i \uparrow}b_{i \downarrow}\Phi_{GS} 
=\tilde{\sum_{\tilde{\sigma}}}
\left[ g(\tilde{\sigma})-g(\tilde{\sigma}_{i\leftrightarrow i-1})\right]
\prod_{j,j\neq i} \tilde{\phi}_{j,\tilde{\sigma}(j)}^{\dagger}|0\rangle ,
\end{eqnarray*}
where $\tilde{\sum}$ stands for a summation over $\tilde{\sigma}$
that satisfies $\tilde{\sigma}(i-1)=\downarrow$ and
$\tilde{\sigma}(i)=\uparrow$. From this we can see that 
$g(\tilde{\sigma})=g(\tilde{\sigma}_{x\leftrightarrow y})$ 
for any $x,y \in \{1\cdots L\}$,
where $\tilde{\sigma}_{x\leftrightarrow y}$ 
denotes the spin configuration 
obtained by exchanging $\tilde{\sigma}(x)$ and $\tilde{\sigma}(y)$. 
The lowest energy state (${\cal H}_U \Phi_{GS}=0$) 
with a given total $S_z (=\frac{1}{2}(L_{\uparrow}-L_{\downarrow}))$ 
is unique because all the $g(\tilde{\sigma})$'s 
for each $S_z$ sector
take the same value and 
factor out.
This is exactly the ferromagnetic ground states, 
since the fully polarized state does not experience $U$, 
and this concludes the proof of the above statement.

In the present model a dispersive band lies below the flat one unlike 
the Mielke's or Tasaki's model.  We can expect that a small enough 
$U$ does not mix the bands 
since they are separated with a finite gap, but this has to be 
examined.  For that purpose 
we take account of the one-electron states 
$\{ \psi_{i,\sigma} \}_{i=1 \cdots L_{<}}$ on the dispersive band 
as well (where the number of states $L_{<}$ is just $L$ here, since we have only one dispersive band below the flat one). 
As Mielke \cite{Mielkeg1} discussed, sufficiently small $U$ can
be treated with a degenerate perturbation theory, and
we can show that the ground state of ${\cal H}={\cal H}_0+{\cal H}_U$
with $L+2L_<$ electrons still has the total spin $S=L/2$ for a small $U$.

By contrast, this approach becomes invalid 
for a large $U$, so that it is necessary to determine 
the total spin of the ground state from other methods.  
We have determined this numerically 
with the exact diagonalization.  
First, let us consider the case of $\Delta \varepsilon=0$.
In fig.\ref{8-16site}, we show the result for the case of 
6 electrons in 8 sites ($L=2$) or 12 electrons in 16 sites ($L=4$), 
for which the band filling is $n=3/4$ with the 
flat band being half-filled.  
We denote the lowest energy among the states
having a given total $S_z$ by $E_{\rm min}(S_z)$, and 
plot 
$\Delta \equiv E_{\rm min}(S_{\rm max})-E_{\rm min}(0)$, 
where $S_{\rm max}=L/2$ is the $S_z$ when the flat band is fully spin 
polarized 
($S_{\rm max}=1$ for the 8-site case, 
$S_{\rm max}=2$ for the 16-site case), as a function of $U$.
In agreement with the above analysis, the ground state has indeed 
the polarized flat band from an infinitesimal $U$.  

Our second new finding is that the ferromagnetism is {\it destroyed} 
when the interaction is too large 
($U>U_C$).  
The critical value $U_C$ 
at which the ground state becomes unpolarized 
is seen to be $U_C =2.3 \sim 2.4$, where 
the sample-size dependence is small.  
To the best of our knowledge this is the first example of 
the ferromagnetism that is unstable for large interactions.

In order to check whether the transition to the unpolarized 
state is abrupt, we plot in 
the inset of fig.\ref{8-16site} 
$\Delta=E_{\rm min}(2)-E_{\rm min}(1)$ 
along with $\Delta=E_{\rm min}(2)-E_{\rm min}(0)$ 
against $U$ for the 16-site system.  
We can see that 
the transition from the state with 
$S=L/2$ directly into 
the $S=0$ state.

So far we have considered the case of half-filled flat band.  
In the As/Si system this would require a doping with 
e.g., alkali metal atoms.\cite{doping}  
The ferromagnetism is expected to be degraded when the 
flat band is pushed away from the half filling.   
It is believed that one-dimensional Tasaki's model is 
ferromagnetic only when the flat band is half-filled, 
while paramagnetic for lower electron densities.
In order to see if this is also the case in the present 
model, we calculate
$\Delta=E_{\rm min}(1)-E_{\rm min}(0)$ as a function of $U$ 
for the number of electrons decreased from 12 to 10 
(quarter-filled flat band) 
in the 16-site system. 
For the parameter region we have investigated, the 
ground state is paramagnetic, i.e., 
the model shares the instability against the hole doping 
with the one dimensional Tasaki's model.

We must also look into how the ferromagnetism proposed here is 
robust against 
the degradation of the flatness of the band.  
For Tasaki's model, Kusakabe and Aoki\cite{Kusakabe1,Kusakabe2},
and later Tasaki\cite{Tasaki2} have shown that
ferromagnetic ground state survives small perturbations 
that make the flat band dispersive.  
Here we look at the stability of the ferromagnetism
against the perturbation,
\[
{\cal H}'= \delta t \sum_{i,\sigma} (c_i^{\dagger} b_{i+1} + {\rm h.c.}).
\]
There are two reasons why we consider such an extra transfer.
First, in real atomic wires, there should be a finite transfer
between the adjacent As clusters, 
and secondly, the ferromagnetism on the 
ladder systems\cite{Kohno} 
is subject to recent investigations, so that the effect of a perturbation 
toward the ladder is of interest.  
In fig.\ref{pert}, we plot
$\Delta=E_{\rm min}(S_{\rm max})-E_{\rm min}(0)$ 
as a function of $U$ for $\delta t=0.1, 0.2$ for 6 electrons in 8 sites.
We can see that the ferromagnetism survives $\delta t = 0.1$, while 
destroyed for $\delta t = 0.2$.  

Next, we look at the robustness of the ferromagnetism 
for the general value of $\Delta\varepsilon\neq 0$ 
($\theta\neq -\pi/6$).  
We can start from an observation 
that the ferromagnetism will be lost for 
$\theta = 0$ or $-\pi/2$, 
since $\{\phi_{i,\sigma}\}$ fails to satisfy the
connectivity condition for $\theta = 0$ invalidating Mielke-Tasaki's
argument, or the lattice becomes disconnected with $t_3=0$ 
for $\theta =-\pi/2$.
We have numerically obtained 
$U_C$ as a function of $\Delta\varepsilon$
for 8 sites in fig.\ref{onsite}. 
We can see that $U_C$ is finite between 
$\theta = 0$ and $-\pi/2$, 
taking its maximum at 
$\Delta\varepsilon\sim 0.5$ ($\theta \simeq -\pi /4$), 
or, equivalently, 
$\Delta\varepsilon\sim -0.5$ with $t_1, t_2 <0$.

Finally let us make a brief comment.
If we change $t_2$ from positive to negative, 
the flat band becomes the lowest band, where 
the ferromagnetic ground state is realized for arbitrary $U>0$ 
when the flat band is half-filled as in Tasaki's model or Mielke's model.

To summarize, we have proposed that the flat-band 
ferromagnetism can occur in a non-frustrated structure that 
can model atomic-scale quantum wires.
Further investigations with the density matrix renormalization
group method will be published elsewhere.
Also, the fabrication of the As/Si(100) wire 
is experimentally under way. 
R.A. is grateful to K. Kusakabe for 
illuminating discussions. Numerical computations were done on
FACOM VPP 500/40 at the Supercomputer Center, Institute
of Solid State Physics, University of Tokyo.
%%%%%%%%%%%%%%%%%%%%  References %%%%%%%%%%%%%%%%%%%%%%%%%%%%%%%%

%%%%%%%%%%%%%%%%%%%% Figure Captions %%%%%%%%%%%%%%%%%%%%%%%%%%%%%%%

\begin{figure}
%\epsfile{file=pic.eps,width=6cm}
\caption{
A tight bindig model proposed by Yajima {\it et al}.  
We have indicated the eigenstate on the flat band.  
Open circles represent As atoms, while filled circles Si.
Shaded area depicts a `Wannier' state.
}
\label{yajimodel}
\end{figure}

\begin{figure}
%\epsfile{file=band.eps,width=8cm}
\caption{
The dispersion relation for the tight-binding model on 
the connected square lattice for a general value of $\theta$ (a) 
and for $\theta = - \pi /6$ ($\Delta \varepsilon=0$) (b).
%$\theta = -\pi /3$ ($\Delta \varepsilon =\sqrt{3} -1$).
}
\label{band}
\end{figure}

\begin{figure}
%\epsfile{file=8site.eps,width=8cm}
\caption{The difference in energy, $\Delta \equiv 
E_{\rm min}(S_{\rm max})-E_{\rm min}(0)$ 
for 6 electrons in 8 sites (squares), and
for 12 electrons in 16 sites (circles).
In the inset, we plot the difference in energy, 
$E_{\rm min}(1)-E_{\rm min}(0)$ (squares) 
and $E_{\rm min}(2)-E_{\rm min}(0)$ (circles) 
for 12 electrons in 16 sites.
}
\label{8-16site}
\end{figure}

\begin{figure}
%\epsfile{file=pert.eps,width=8cm}
\caption{
A similar plot as in Fig. 3 for 
$\Delta=E_{\rm min}(S_{\rm max})-E_{\rm min}(0)$ when 
an extra transfer $\delta t (=0.1$:circles or 0.2:squares) 
is introduced (as in the inset) for 6 electrons in 8 sites.
}
\label{pert}
\end{figure}

\begin{figure}
%\epsfile{file=onsite.eps,width=8cm}
\caption{
$U_C$ as a function of $\Delta\varepsilon$
for 6 electrons in 8 sites.
}
\label{onsite}
\end{figure}

\end{multicols}
\end{document}